\documentclass[twocolumn,secnumarabic,amssymb, nobibnotes, aps]{revtex4-2}
\usepackage{graphicx}

\begin{document}

\title{Optical phased array neural probes for beam-steering in brain tissue}

\author{Wesley D. Sacher$^{1,2,3,\dagger,*}$}
\author{Fu-Der Chen$^{2,3,\dagger}$}
\author{Homeira Moradi-Chameh$^4$}
\author{Xinyu Liu$^1$}
\author{Ilan Felts Almog$^2$}
\author{Thomas Lordello$^2$}
\author{Michael Chang$^4$}
\author{Azadeh Naderian$^4$}
\author{Trevor M. Fowler$^1$}
\author{Eran Segev$^1$}
\author{Tianyuan Xue$^2$}
\author{Sara Mahallati$^4$}
\author{Taufik A. Valiante$^{4,5,6}$}
\author{Laurent C. Moreaux$^1$}
\author{Joyce K. S. Poon$^{2,3}$}
\author{Michael L. Roukes$^1$}

\affiliation{$^1$Division of Physics, Mathematics, and Astronomy, California Institute of Technology, Pasadena, California 91125, USA}
\affiliation{$^2$Department of Electrical and Computer Engineering, University of Toronto, 10 King's College Rd., Toronto, Ontario M5S 3G4, Canada}
\affiliation{$^3$Max Planck Institute of Microstructure Physics, Weinberg 2, 06120, Halle, Germany}
\affiliation{$^4$Krembil Research Institute, Division of Clinical and Computational Neuroscience, University Health Network, Toronto, Ontario, Canada}
\affiliation{$^5$Division of Neurosurgery, Department of Surgery, Toronto Western Hospital, University of Toronto, Toronto, Ontario, Canada}
\affiliation{$^6$Institute of Biomaterials and Biomedical Engineering, University of Toronto, Toronto, Ontario, Canada}
\affiliation{$^\dagger$Equal contribution}
\affiliation{$^*$Corresponding author: wesley.sacher@mpi-halle.mpg.de}

\begin{abstract}
Implantable silicon neural probes with integrated nanophotonic waveguides can deliver patterned dynamic illumination into brain tissue at depth. Here, we introduce neural probes with integrated optical phased arrays and demonstrate optical beam steering \emph{in vitro}. 
Beam formation in brain tissue was simulated and characterized. The probes were used for optogenetic stimulation and calcium imaging.
\end{abstract}

\maketitle


Genetically encoded optogenetic actuators and fluorescence indicators have become powerful tools in the interrogation of brain activity, since they enable the control and imaging of neurons  with high cell-type specificity and single-cell spatial resolution \cite{DeisserothNatNeuro2015, DanaPLOSOne2014,VILLETTECell2019}. Today's optical systems for optogenetics and functional fluorescence imaging, such as multi-photon microscopes and implantable fiber optics, are typically built from bulk off-the-shelf components and are physically large and complex \cite{PackerNatureMethods2015}. Yet, advances in silicon (Si) integrated photonics have led to the dense integration of nanoscale waveguides and devices into millimeter-scale circuits that achieve complex functions \cite{SunNature2013,DoerrFrontiersInPhysics2015}. Thus, Si photonic technology can be leveraged to create nanophotonic tools that miniaturize optical systems for neurobiology and deliver light into brain tissues in ways that are not possible with bulk optics. One approach is to realize implantable chip-scale photonic devices that deliver and control patterned illumination in brain tissues at depths inaccessible by free-space optics, i.e., beyond the optical attenuation length. Along these lines, nanophotonic waveguides with grating coupler (GC) light emitters \cite{SegevNeurophotonics2016,LibbrechtJournalNeurophysiology2018,Sacher_Neurophotonics_2021,MohantyNatureBiomedicalEngineering2020} and micro-light-emitting-diodes ($\mu$LEDs) \cite{WuNeuron2015} have been integrated onto implantable Si probes. In brain tissues, since light mostly scatters forward \cite{YonaENEURO2015}, low-divergence beams can be emitted from GCs over distances of 200-300 $\mu$m \cite{SegevNeurophotonics2016,LibbrechtJournalNeurophysiology2018}. Compared to $\mu$LEDs, nanophotonic waveguide-based probes do not generate excess heat beyond that caused by the light itself, can more precisely tailor the optical emission profile, are compatible with wafer-scale foundry manufacturing \cite{SacherOE2019, Sacher_Neurophotonics_2021}, and can achieve a high light source density. Furthermore, as evidenced by the recent advancements in Si photonic beam-forming \cite{SunNature2013,PoultonOL2017,HutchisonOptica2016}, sophisticated gratings and photonic circuit designs can enable precisely patterned illumination with high spatial resolution. 

Here, we report the first implantable Si neural probes capable of optical beam-steering in tissue.  The probes use silicon nitride (SiN) optical phased arrays (OPAs) as light emitters; the emitted beams were steered by wavelength tuning. The OPAs were designed to operate at blue wavelengths for the excitation of the opsin Channelrhodopsin-2 (ChR2) and the genetically encoded calcium indicator GCaMP6. The probes were validated \emph{in vitro} in mouse brain slices, demonstrating sufficient power for optogenetic stimulation and functional imaging as well as spatial control of the beam on the neuron scale. A preliminary report of this work appeared in \cite{SacherCLEO2019_beam_steering}.

\begin{figure*}
    \centering
    \includegraphics[width= 6.4in]{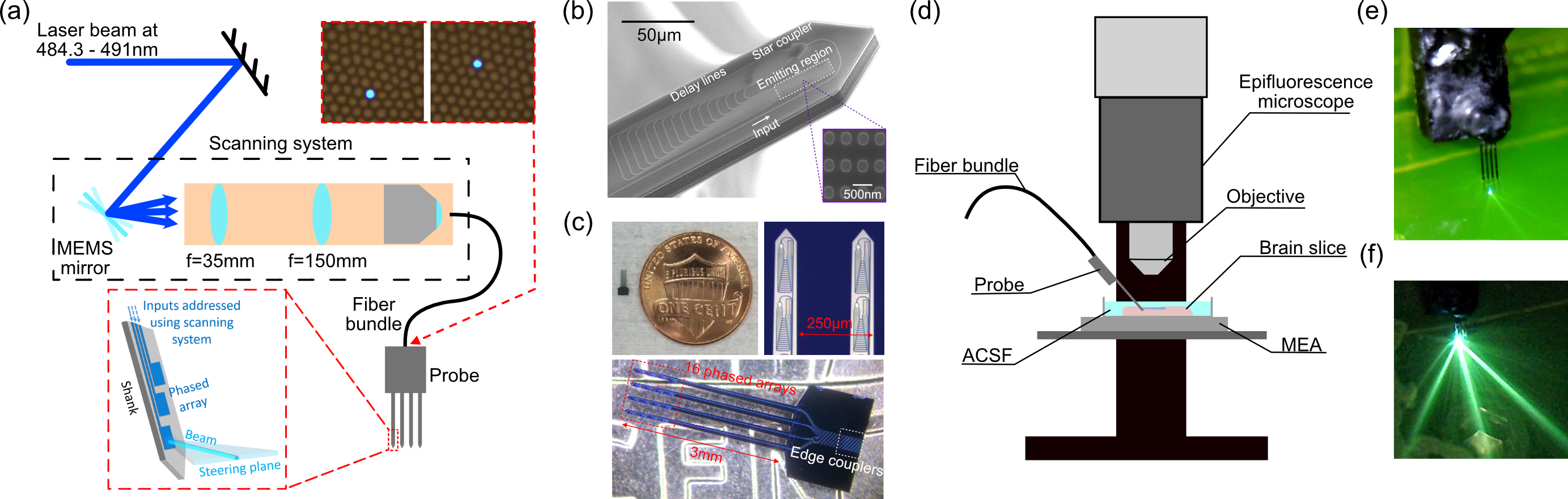}
    \caption{OPA neural probes. (a) Schematic of the OPA neural probe connected to the scanning system (inset) optical micrographs of an image fiber bundle facet with cores addressed by the scanning system. (b) Scanning electron micrograph (SEM) of one of the shanks of an OPA neural probe (inset) SEM of a portion of the SiN gratings prior to top cladding deposition during fabrication. (c) Photograph and micrographs of an OPA neural probe. (d) Illustration of the experimental apparatus for testing the neural probes in brain slices. Abbreviations: microelectrode array (MEA), artificial cerebrospinal fluid (ACSF). (e, f) Photographs of the packaged OPA neural probe emitting light in fluorescein; (e) has a smaller field of view and external illumination applied for visibility of the shanks, (f) shows the emitted beams (zoomed out, no illumination). (b, c, e, f) are from \cite{SacherCLEO2019_beam_steering}; the fiber bundle inset of (a) is from \cite{Sacher_Neurophotonics_2021}.}
    \label{fig:Figure1}
\end{figure*}

Figure \ref{fig:Figure1} shows the OPA neural probes, which consisted of 4 shanks, each $\sim 18$ $\mu m$ thick, 3 mm long, and 50 $\mu m$ wide, on a 250 $\mu m$ pitch, and a thicker base region. The SiN waveguides were 200 nm thick. The probes were fabricated on 100 mm diameter silicon-on-insulator wafers as described in \cite{SegevNeurophotonics2016}. On each shank were 4 OPAs, with the design shown in Fig. \ref{fig:Figure1}(b). A star coupler split the light in the input waveguide into 16 delay line waveguides. The delay lines were routed for a differential path-length, and each terminated with a light-emitting grating. As the input wavelength was tuned, the differential phase-shift between the light-emitting gratings led to the angular steering of the emission \cite{VanAcoleyenPTL2011}. The delay lines consisted of waveguides with an initial single-mode width of 240 nm following the star coupler for a length of 12 $\mu$m that adiabatically widened to 400 nm to reduce phase error. The differential length of the delay lines was 16 $\mu$m, chosen so the free spectral range (FSR) ($\sim 6$ nm) matched the wavelength tuning range of our external cavity diode laser of 484.3 to 491 nm. The pitch and width of the arrayed gratings in the OPA were 700 nm and 300 nm, respectively. The period of each grating was 440 nm; in water, the steering plane [Fig. \ref{fig:Figure1}(a)] was angled at $\sim 25^{\circ}$ from the normal of the probe. The OPAs were designed for low crosstalk between the arrayed waveguides. The array pitch was significantly larger than the half-wavelength criterion for emission of a single grating order, and typically 3 lobes were emitted from each OPA. Optimization of the array pitch, increased SiN thickness for higher optical confinement and lower crosstalk, and apodization \cite{HutchisonOptica2016} may suppress these additional lobes. In the following, we focus on one of the OPA neural probes that was packaged and studied in detail.

\begin{figure}[!b]
    \centering
    \includegraphics[width=3.1in]{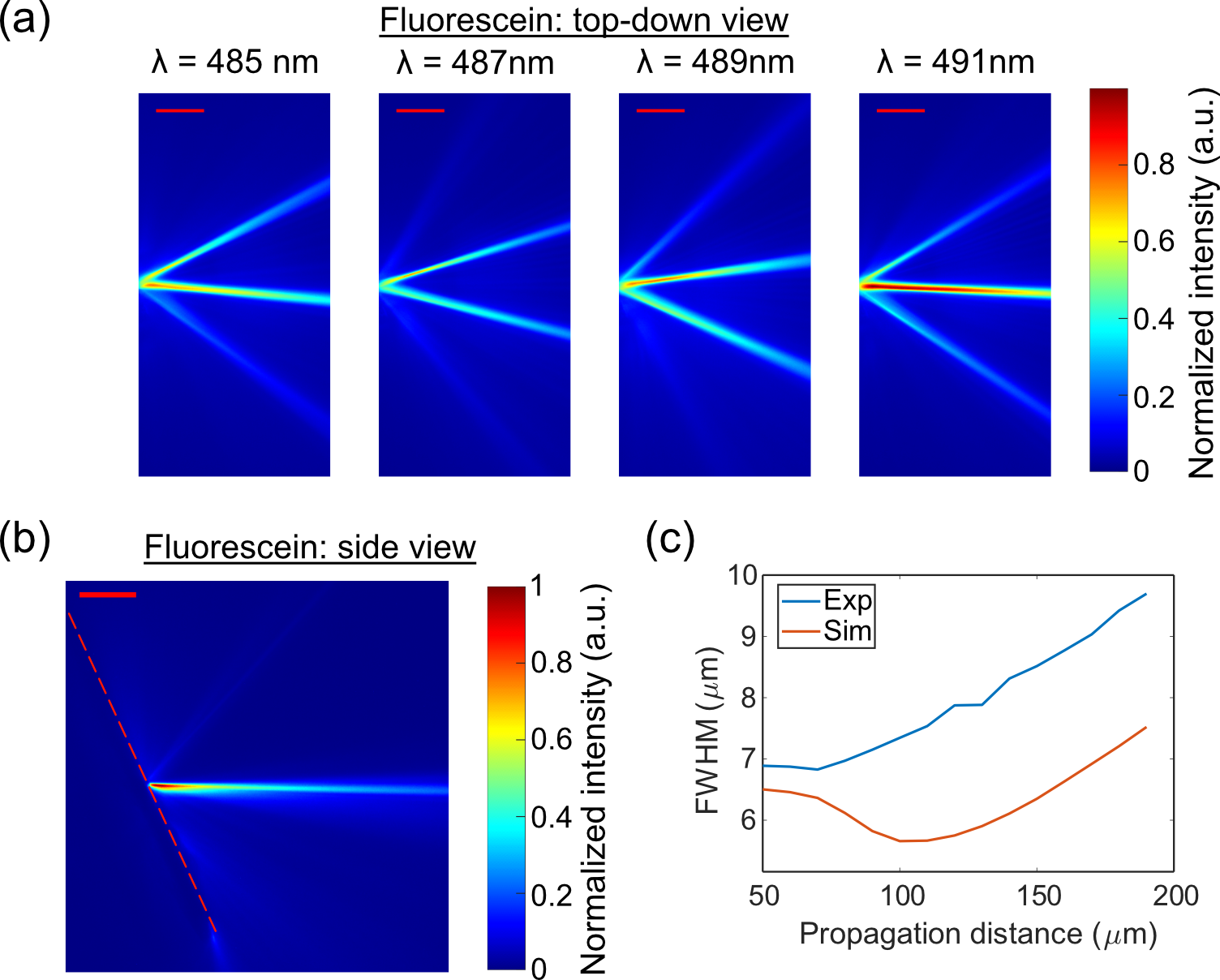}
    \caption{Characterization of the neural probe beam profiles in fluorescein. (a) Top-down beam profiles at various wavelengths ($\lambda$). (b) Side profile of the beam at $\lambda = 484.3$ nm; the top surface of the shanks is delineated by the dashed line. The scale bars are (a) 50 $\mu$m, (b) 100 $\mu$m. (c) Top-down measured (``Exp'') and simulated (``Sim'' in water) FWHM beam width versus propagation distance of the central lobe at $\lambda = 491$ nm.}
    \label{fig:Figure2}
\end{figure}

The probe was passive to reduce tissue heating, and the 16 OPAs on the probe were independently addressed using the spatial addressing scheme in \cite{ZorzosOL2012,Sacher_Neurophotonics_2021}.  As shown in Fig. \ref{fig:Figure1}(a), a micro-electro-mechanical system (MEMS) mirror deflected a laser beam into individual cores of an image fiber bundle, which was attached to the probe base.  Each fiber core was aligned to an edge coupler on the probe, which was connected to an OPA on the shank. The switching time of the MEMS mirror was $\sim 5$ ms. About $\sim 10$ $\mu W$ was emitted from each OPA.

\begin{figure*}[!ht]
    \centering
    \includegraphics[width= 6.0in]{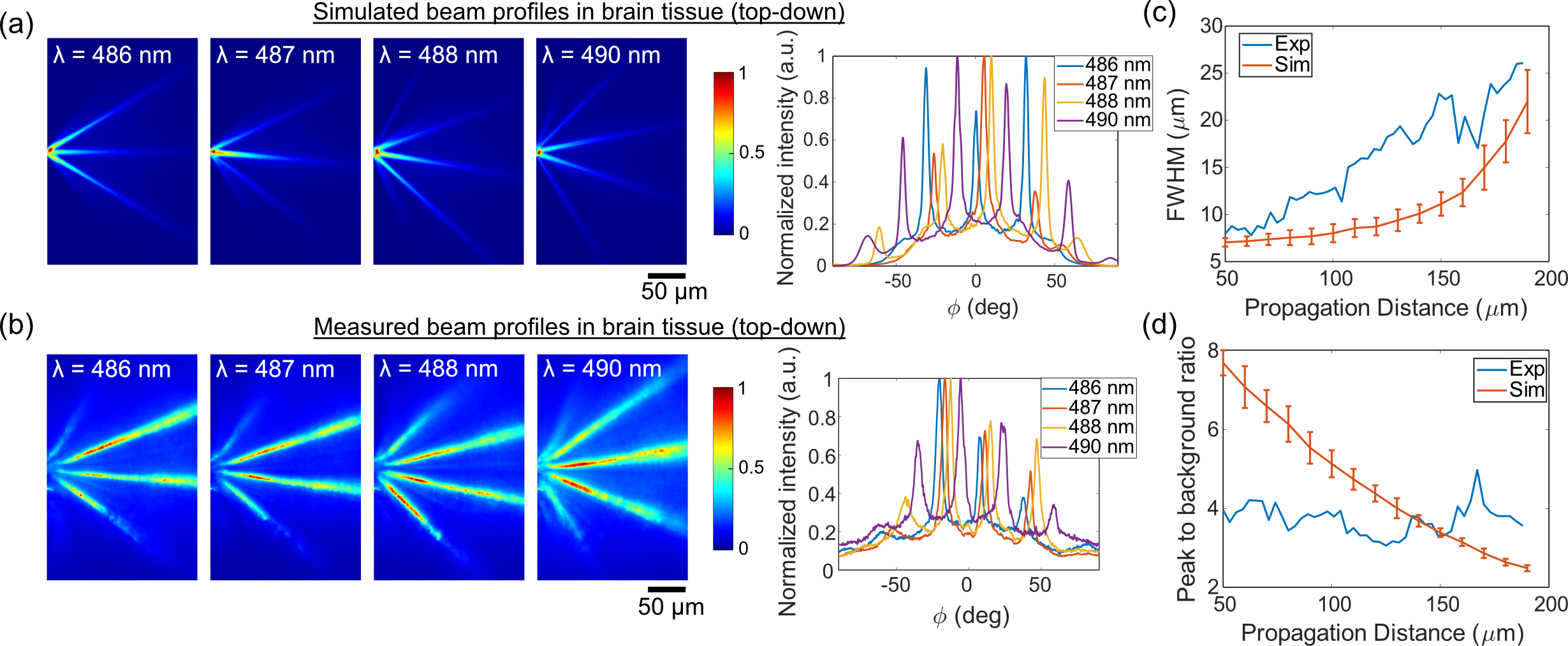}
    \caption{OPA intensity top-down beam profiles in brain tissue. (a) Simulated and (b) measured beam profiles at 4 wavelengths. In (b), the OPA neural probe was inserted into a brain slice from a VGAT-ChR2-EYFP mouse [see Fig. \ref{fig:Figure1}(d)]. Left: top-down intensity beam profile images; right: intensity versus emission angle ($\phi$) at a propagation distance of 100 $\mu$m (i.e., along arcs of radius 100 $\mu$m centered on the emitting region of the OPA). (c) Measured and simulated FWHM beam width versus propagation distance. (d) Measured and simulated ratios of peak beam intensity to background intensity versus propagation distance. The wavelength in (c,d) is 490 nm. The simulated results in (c,d) are an average over 10 simulations with different phase masks (see Supplementary Materials) and the bars represent the standard deviation of the 10 simulations; the beam profiles in (a) are from a single simulation.}
    \label{fig:Figure3}
\end{figure*}

Following the method in \cite{Sacher_Neurophotonics_2021}, to characterize the emitted beam profiles in non-scattering media, the probe was immersed in a fluorescein solution and the fluorescence was imaged. The setup in Fig. \ref{fig:Figure1}(d) was used, but with the chamber replaced with a container of fluorescein; the probe was angled so the emitted beams were parallel to the fluid surface. Top-down images of the emission pattern from the probe at various wavelengths are shown in Fig. \ref{fig:Figure2}(a). Over a wavelength tuning range from 484.3 to 491 nm, the beams were steered continuously $\pm 16^{\mathrm{o}}$ with narrow beams formed within a distance of 300 $\mu m$ (Fig. \ref{fig:Figure2}). The two side lobes were at angles of about $\pm 32^\mathrm{o}$ from the main beam. Figure \ref{fig:Figure2}(b) shows the beam profile imaged from the side; the full-width-at-half-maximum (FWHM) thickness was $< 19$ $\mu m$ over a 300 $\mu m$ propagation distance. The peak intensity of the main lobe was 7 to 17$\times$ larger than the background light intensity at propagation distances of 50 to 300 $\mu$m, Fig. \ref{fig:Figure2}(a). A significant component of the background was due to optical scattering from the photonic circuit.

OPA beam formation in brain tissue was verified and investigated in simulation and experiment. The simulation method is described in Supplementary Materials; the scattering coefficient ($\mu_s$) $= 200$ cm$^{-1}$, the absorption coefficient ($\mu_a$) $= 0.62$ cm$^{-1}$, and the anisotropy of tissue ($g$) $= 0.83$. The simulated beam profiles, Fig. \ref{fig:Figure3}(a), show that the optical scattering leads to broadening of the emitted beams, a background generated between the lobes, and a reduction in intensity with propagation distance. Over the majority of the steering range, the FWHM beam width is $< 17$ $\mu$m and the ratio of peak beam intensity to background intensity is $> 2.8$ for propagation distances of 50 to 150 $\mu$m, Figs. \ref{fig:Figure3}(c), \ref{fig:Figure3}(d), S1, and S2. Experimental validation of the OPAs in brain tissue was performed by inserting the probe at shallow depths (OPA $<100$ $\mu$m from the surface) into perfused \emph{in vitro} brain slices from adult transgenic mice co-expressing Channelrhodopsin-2 and yellow fluorescent protein (VGAT-ChR2-EYFP, \cite{ZhaoNatureMethods2011}) and another strain of mice expressing the genetically-encoded calcium indicator GCaMP6s (Thy1-GCaMP6s, \cite{DanaPLOSOne2014}). The tissue preparation is described in Supplementary Materials. All experimental procedures described here were reviewed and approved by the animal care committees of the University Health Network in accordance with the guidelines of the Canadian Council on Animal Care.

\begin{figure}[!b]
    \centering
    \includegraphics[width= 3.35in]{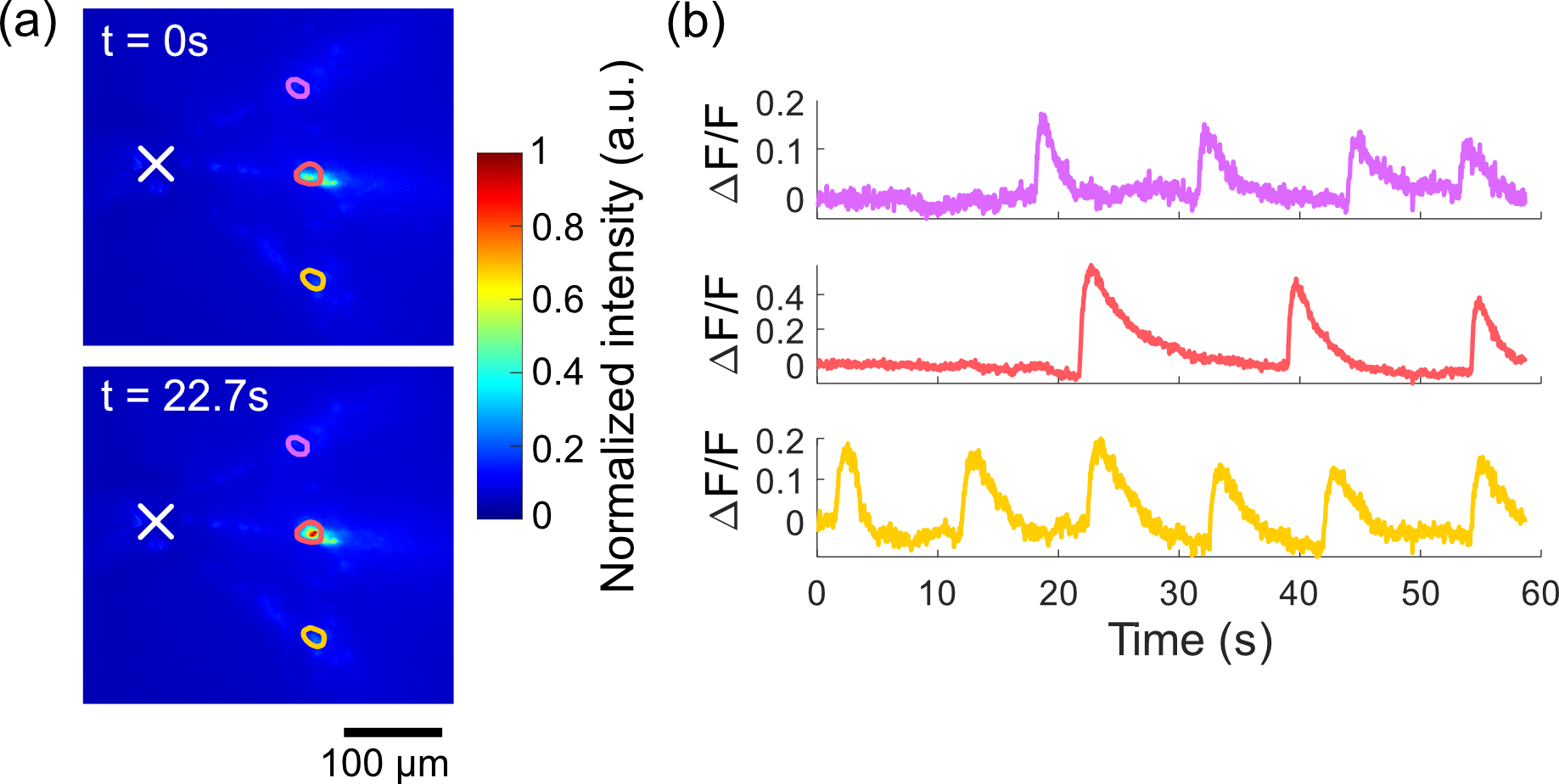}
    \caption{Functional imaging test of the OPA neural probe inserted into a brain slice from a Thy1-GCaMP6s mouse. (a) Top-down fluorescence images at times t = 0 s and 22.7 s showing a calcium event in the central lobe; the OPA emitter position is marked with ``X''. (b) Change in fluorescence ($\Delta F/F$) time traces of the regions of interest delineated in (a).}
    \label{fig:Figure4}
\end{figure}

Fluorescence images of the OPA beam profiles in cerebellar brain slices from the VGAT-ChR2-EYFP mice were captured using the excitation of the YFP by the OPA illumination and the setup in Fig. \ref{fig:Figure1}(d). Steerable beams were formed, as shown in Figs. \ref{fig:Figure3}(b)-(d) and S2, with the FWHM beam width $< 23$ $\mu$m and the ratio of peak beam intensity to background intensity $>2.7$ for propagation distances of 50 to 150 $\mu$m. Cerebellar slices were selected for their relatively uniform YFP labeling; some non-uniformity was evident from the intensity maxima in the Fig. \ref{fig:Figure3}(b) top-down images. As a simple test of functional imaging, the probe was inserted into a hippocampal brain slice from a Thy1-GCaMP6s mouse. The lobes of the OPA illuminated 3 neurons, or small clusters thereof, and spontaneous time-dependent fluorescence was detected with high contrast, Fig. \ref{fig:Figure4}. The rise and fall times of the calcium events agreed with reported GCaMP6s dynamics \cite{DanaPLOSOne2014}.

\begin{figure*}
    \centering
    \includegraphics[width=\textwidth]{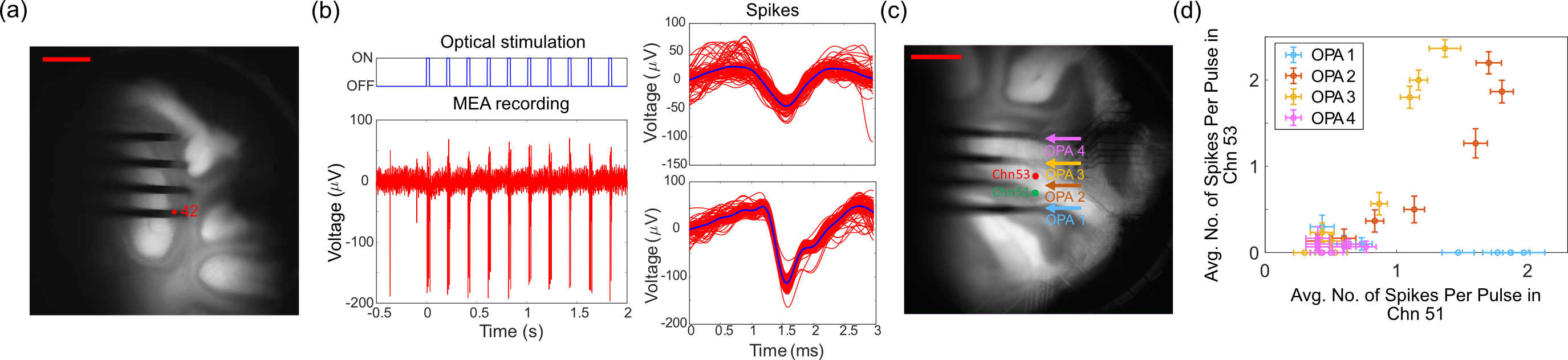}
    \caption{Optogenetic stimulation tests of the OPA neural probe (fixed wavelength) inserted into brain slices on an MEA. (a,b) Illumination was applied from a single OPA, and (c,d), in a different brain slice, illumination was applied sequentially from 4 OPAs. The fluorescence images (a,c) show the relative positions of the OPAs and the relevant MEA electrodes for (b,d); epi-illumination from the top-down microscope was applied for visibility of the shanks and the structure of the brain slice via the YFP labeling. The scale bars are 500 $\mu$m. (b) Extracellular recording and examples of sorted spikes from Channel (``Chn'') 42 of the MEA, as marked in (a), with illumination from an OPA. (d) Extracellular recording showing the average number of spikes per optical pulse from 4 of the OPAs on MEA Chn 53 and 51 as marked in (c). Each point is from one stimulation trial consisting of 3 sets of 10 optical pulses; the pulse width was 50 ms, the period was 200 ms, and the recovery period between each set of 10 pulses was 10 s. The optical power was reduced across the trials for each OPA. The error bars represent the standard error of the mean spikes per pulse during a trial.}
    \label{fig:Figure5}
\end{figure*}

Optogenetic stimulation was tested by inserting the probe into VGAT-ChR2-EYFP mouse cerebellar brain slices. Here, the probe orientation was modified compared to Fig. \ref{fig:Figure1}(d) so that the probe emitted light nearly orthogonal to the tissue surface (see Fig. S3, Supplementary Materials). Pulsed light was applied, and extracellular electrophysiological signals were recorded by a microelectrode array (MEA) beneath the brain slice. Robust spiking with illumination from the neural probe was observed, Fig. \ref{fig:Figure5}(b), showing that the optical intensity was sufficient for optogenetic stimulation. However, the precise positions of the stimulated neurons relative to the OPA were unknown. Spatial differences in the activity patterns across the MEA were not always repeatable with the steering of the beam. We hypothesize this was due to one or more of the following reasons: 1) the interconnectedness of the neurons, 2) ChR2 expression in inhibitory rather than excitatory cells in VGAT-ChR2-EYFP mice, and 3) the spacing between the OPA lobes and background between them. Figure \ref{fig:Figure5}(d) supports the first 2 hypotheses. Figure \ref{fig:Figure5}(d) shows the correlation in the spiking activity recorded at two microelectrodes in the same column (Channels 53 and 51) in the MEA as a result of the excitation from 4 of the OPAs on the probe. Here, illumination from one OPA on each shank was sequentially applied (the OPAs belong to the same row), and a number of trials with decreasing optical power were performed. The spike rates on Channels 53 and 51 indicated that lateral shifts of the illumination on the order of the shank pitch (250 $\mu$m) were necessary for repeatable changes in the activity patterns across the MEA (e.g., comparing the trends of OPAs 1 and 2). Generally, the number of spikes per pulse increased with the optical power. The experiments, data analysis, and additional data (Fig. S4) are detailed in Supplementary Materials. 

In summary, we have presented the first implantable Si neural probes with integrated OPAs for generating steerable optical beams in brain tissue. Beam formation was confirmed in simulation and \emph{in vitro} brain slices, and the emitted intensity was sufficient for optogenetic stimulation and functional imaging. The OPA feature sizes are compatible with photonic foundry processes \cite{SacherOE2019,Sacher_Neurophotonics_2021}; OPAs at green/red wavelengths can be realized with larger feature sizes for excitation of other opsins and fluorophores. These results show the possibilities for delivering patterned, dynamic illumination at depth in brain tissue for brain activity mapping \cite{MoreauxNeuron2020}.

\medskip

\noindent \textbf{Funding.} National Institutes of Health (NS090596, NS099726); Caltech Kavli Nanoscience Institute; Canadian Institutes of Health Research; Natural Sciences and Engineering Research Council of Canada; Canadian Foundation for Innovation.

\medskip

\noindent \textbf{Disclosures.} The authors declare no conflicts of interest.

\pagebreak
\widetext
\begin{center}

\textbf{\large Optical phased array neural probes for beam-steering in brain tissue:\\Supplementary materials}
\end{center}
\setcounter{equation}{0}
\setcounter{figure}{0}
\setcounter{table}{0}
\setcounter{page}{1}
\makeatletter
\renewcommand{\theequation}{S\arabic{equation}}
\renewcommand{\thefigure}{S\arabic{figure}}
\renewcommand{\bibnumfmt}[1]{[S#1]}
\renewcommand{\citenumfont}[1]{S#1}

\section{Optical Phased Array Beam Profile Simulations}

We applied the beam propagation method (BPM) to simulate the optical scattering in tissue  \cite{cheng2019development}. The model simulates the forward propagation of a beam in tissue with an iterative two-step process: 1) a two-dimensional phase mask is applied to the propagating beam, the phase mask has a small phase variance about the mean phase, which is equal to 0, and 2) beam propagation between the phase masks is assumed to only diffract, and it is modeled with the angular spectrum method \cite{goodman2005introduction}. The scalar electric field, $E$, is given by
\begin{equation}
E(k_x,k_y,z+d)=E(k_x,k_y,z) \mathrm{e}^{in\sqrt{k^2-k_x^2+k_y^2}d},
\end{equation}
where $d$ is the propagation distance, $n$ is the average refractive index of the medium, and $k$ is the wavenumber ($k_x$ and $k_y$ are  $x$ and $y$ components of the corresponding wavevector). The amplitude of any wave with $k_x^2+k_y^2>k$ is set to 0 as it is evanescent. The model is applicable to  brain tissue since the beam is mostly forward scattered and its simulation results have been validated with the analytical solutions from the radiative transfer equation \cite{cheng2019development}.

\par

The patterns of the phase masks determine the scattering properties of the medium. We followed the design strategy in \cite{cheng2019development} to create the phase mask patterns. Each phase mask introduced a spatially varying random phase to the fields with the statistical properties governed by $\sigma_p$ and $\sigma_x$. $\sigma_p$ adjusted the variance of the phase introduced by each phase mask, which affected the scattering coefficient. Each phase mask was smoothened by a Gaussian filter with variance $\sigma_x$ to control the phase correlation between pixels, which affected the anisotropy value. We also multiplied the beam profile with the attenuation factor after each propagation step. For the scattering simulation, we set the scattering coefficient ($\mu_s$) to 200 $cm^{-1}$, the anisotropy of the tissue ($g$) to 0.83, and the attenuation coefficient ($\mu_a$) to 0.62 $cm^{-1}$ \cite{yona2016realistic}.

\par

The field 3 $\mu m$ above the optical phased array (OPA) was calculated and used as the launch field for the above propagation simulations. The SiO$_2$ cladding thickness above the OPA was 1 $\mu m$. To calculate the launch field, the emission field from a single grating emitter array element of the OPA was first obtained (3 $\mu m$ above the grating) using finite difference time domain (FDTD) simulations, with the simulations performed for transverse magnetic (TM) polarized light in the waveguides. Then an array of single grating emitter field profiles was generated via laterally shifting the position of each subsequent emitter by the array pitch and applying phase shifts to each field profile corresponding to the wavelength and delay line parameters of the array element. Finally, the OPA field was the sum of the arrayed field profiles. This process was repeated for each simulated wavelength. The nominal OPA dimensions reported in the manuscript were used for the simulations. In addition, we corrected for the amplitude distribution across the output ports of the OPA star coupler by multiplying the emitter field profiles by an envelope function. We set the simulation domain volume to be $500 \mu m$ x $500 \mu m$ x $250 \mu m$ to avoid any edge effects while containing the entire simulation in the computer memory.

\par
The same simulation model was used to study the OPA beam profiles in a non-scattering medium (water) with all phase masks set to a uniform value of 0. For verification, we compared cross-sections of beam profiles obtained from the BPM simulations and the conventional diffraction integral method. No significant differences were observed between the two methods at propagation distances of up to 200 $\mu m$.

After the BPM simulations, we performed a series of steps to determine the 2D beam profile that would be expected from a top-down microscope as in the experimental apparatus in Fig. 1(d). First, we stacked the beam profile cross-sections to form a 3D model of the beam in tissue. We then rotated the model so that the beam propagation axis was parallel with the horizontal plane. Bicubic spline interpolation was used to interpolate the pixel values on the new grid of the rotated coordinate system. Lastly, a 2D beam profile was obtained by filtering and adding together the transverse planes of the 3D beam profile model, i.e., all planes parallel to the microscope focal plane. To approximately emulate the microscope resolution limit, we applied a Gaussian filter to each transverse plane with filter size equal to the Gaussian propagation beam waist,
\begin{equation}
w(z)=w_0\sqrt{1+\left(\frac{\lambda z}{\pi w_0^2}\right)^2},
\end{equation}
where $w_0$ is the point spread function of the objective used in the experimental apparatus [Fig. 1(d)], and $z$ is the distance of the transverse plane from the focal plane. For each simulation, the transverse plane with the highest intensity at the input to the scattering medium was selected as the focal plane.

The simulated and measured top-down beam profile FWHM values and peak-to-background ratios in Figs. 2, 3, and S2 were calculated along concentric arcs centered on the OPA emitting region; the radius of each arc was equal to the propagation distance. For FWHM beam width calculations, the maximum was simply the maximum intensity along the arc, while for peak-to-background ratio calculations, to reduce the impact of noise, the peak intensity was defined as the average of the top 1\% of intensity data points along the arc. The background intensity for the peak-to-background ratio calculations was defined as the intensity in the troughs between adjacent lobes. As the OPA beam profiles typically had 3 lobes (Figs. 2 - 3), the background intensity was calculated considering only the higher of the 2 troughs between the 3 lobes. The trough intensity was calculated as the average of the lowest 1\% of intensity data points along the arc in the trough. The arcs spanned the central lobe and the two adjacent troughs.

\section{Experimental apparatus}

The scanning system in Fig. 1(a) is detailed in \cite{Sacher2020}, with the exception that a wavelength-tunable laser was used here rather than a fixed wavelength laser. Briefly, light from the wavelength-tunable laser (TOPTICA Photonics Inc., DLC DL pro tunable laser system with integrated optical isolator, fiber coupler, motorized wavelength tuning, 484.3 - 491 nm wavelength tuning range) was coupled into a single-mode fiber (460-HP, Nufern Inc.), which was connected to an inline fiber polarization controller. The laser light was launched into free space using a fiber collimator, and this free-space laser beam was gated by a mechanical shutter, directed through a variable neutral density filter (for control of optical power), and input into the scanning system, as shown in Fig. 1(a). The scanning system included a 2-axis micro-electro-mechanical system (MEMS) mirror (A7B2.1-3600AL, Mirrorcle Technologies Inc.), two biconvex lenses with 35- and 150-mm focal lengths, and a 20$\times$ objective lens (Plan Apochromat, 20-mm working distance, 0.42 numerical aperture, Mitutoyo Corporation). Actuation of the MEMS mirror enabled addressing of individual cores of the image fiber bundle (Fujikura FIGH-06-300S). The fiber bundle was optically coupled to and packaged together with the OPA neural probe using the method described in \cite{Sacher2020}. The packaged probe was attached to a 4-axis micro-manipulator (QUAD, Sutter Instrument Company) for immersing the probe in the fluorescein solution (10 $\mu$M concentration, pH $> 9$) and inserting the probe into the brain slices. Since the OPA beam profiles were polarization-dependent, the fiber bundle was fixed in position during the experiments (to avoid polarization fluctuations due to movement of the fiber bundle). The input light to the neural probe chip was TM-polarized. 

The maximum optical power available at the input to the scanning system was about 2 mW, and the loss of the scanning system (measured from the free-space input of the scanning system to the distal facet of the fiber bundle) was typically 40 - 60$\%$ (with the neutral density filter set to its minimum loss). The transmission of the OPA neural probe chip (from the facet of an optimally aligned single-mode fiber to the free-space OPA output) was typically about -20 dB. As described in \cite{Sacher2020}, the edge couplers accounted for roughly 10 dB of this loss, and improved edge coupler designs may greatly improve the optical transmission of OPA neural probes. In addition, deviations of the fiber bundle core positions from a regular pitch and fiber misalignment during the optical packaging procedure resulted in significant variations in the coupling efficiency between the fiber bundle cores and the edge couplers of the neural probe. As a result, the total transmission of the OPAs (from the input to the scanning system to the free-space OPA output) varied from about -23 dB to -40 dB, with the majority of the OPAs having transmissions between about -23 to -33 dB.

The experimental apparatus used for OPA beam characterization and \emph{in vitro} testing of the neural probe [Figs. 1(d) and S3] included a Nikon Eclipse FN1 upright epifluorescence microscope with an sCMOS camera (Zyla 4.2 PLUS, Andor Technology Ltd.) and a 10$\times$ objective lens (Plan Apochromat, 34-mm working distance, 0.28 numerical aperture, Mitutoyo Corporation). The images captured by the microscope were inverted (i.e., vertically mirrored). No image processing to correct the image inversion was applied, and the beam profile images [Figs. 2(a), 3(b), and 4(a)] and epifluorescence brain slice images [Figs. 5(a), 5(c), and S4(a)] are inverted. For the fluorescein and Thy1-GCaMP6s mouse brain slice imaging, an EGFP filter cube (49002, Chroma Technology Corporation) was used in the epifluorescence microscope, and for the VGAT-ChR2-EYFP mouse brain slice imaging, an EYFP filter cube (49003, Chroma Technology Corporation) was used. The camera exposure time was: 50 ms for the fluorescein beam profile images in Fig. 2(a), 500 ms for the \emph{in vitro} brain slice beam profile images in Fig. 3(b), and 25 ms for the \emph{in vitro} brain slice calcium imaging in Fig. 4. The microelectrode array (MEA) was a perforated design with 60 titanium nitride electrodes, 30 $\mu m$ electrode diameter, and 100 $\mu m$ electrode pitch (60pMEA 100/30iR-Ti-pr-6 mm high plastic ring, Multi-Channel Systems). MEA electrical activity recordings were performed using a MEA-1060-Up-BC amplifier and the MC Rack software (Multi-Channel Systems). The sampling rate for the MEA recordings was 25 kHz.

\section{Brain slice preparation}

All experimental procedures described here were reviewed and approved by the animal care committees of the University Health Network in accordance with the guidelines of the Canadian Council on Animal Care. Brain slices were prepared from 40 - 80 days old VGAT-ChR2-EYFP (The Jackson Laboratory, stock number 014548) and Thy1-GCaMP6s (The Jackson Laboratory, stock number 025776) mice for the \emph{in vitro} beam profile/optogenetic stimulation and the \emph{in vitro} calcium imaging experiments, respectively; the brain slice preparation is detailed in \cite{Sacher2020}. 350 $\mu$m thick sagittal slices from the cerebellum were used for the \emph{in vitro} beam profile and optogenetic stimulation experiments (Figs. 3 and 5). A 300 - 450 $\mu m$ thick horizontal slice from hippocampus was also tested during the beam profile experiments, but the labeling was significantly more non-uniform compared to the slices from the cerebellum. The calcium imaging experiment used a 350 - 450 $\mu m$ thick horizontal slice from hippocampus. For the optogenetic stimulation and imaging experiments, brain slices were transferred to the MEA chamber, Figs. 1(d) and S3, and perfused with a constant flow of rodent artificial cerebrospinal fluid (ACSF) \cite{TingSciRep2018}, which was continuously aerated with carbogen. During imaging of the Thy1-GCaMP6s mouse brain slice, KCl was added to the ACSF to increase the excitability of the neurons and the amount of spontaneous neuronal activity; the KCl concentration in the ACSF was 30 mM.

\section{Optogenetic Stimulation Protocol}

For the single-OPA optogenetic stimulation experiment shown in Fig. 5(b), 10 optical pulses with a pulse width of 30 ms and a period of 200 ms were applied to the brain slice. A recovery period of 10 s was used after each stimulation pulse train. We repeated the pulse train 10 times.

For the multi-OPA optogenetic stimulation experiments shown in Figs. 5(d) and S4, multiple ``stimulation trials'' were performed, and for each trial, illumination was applied to the brain slice from one of the 4 OPAs in Figs. 5(c) and S4(a). Each trial corresponds to a point in Figs. 5(d) or S4(b), and the optical power was varied between trials. Each stimulation trial consisted of 3 sets of 10 optical pulses; the optical pulse width was 50 ms, the period was 200 ms, and the recovery period between each set of 10 optical pulses was 10 s. 7-8 trials with different optical power settings were performed for each OPA by using the variable neutral density filter in the experimental apparatus to reduce the input optical power to the scanning system in steps of approximately 10$\%$ of the maximum power. The trials were performed in the following order: 1) the optical power was set to the maximum value, 2) a trial was performed for each OPA sequentially (from OPA 1 to 4), 3) the optical power was reduced by a 10$\%$ increment, 4) the next set of OPA trials was performed. This process was repeated until 7-8 trials were performed for each OPA; at the lowest power setting, $< 1$ spike per optical pulse (on average) was observed. This procedure ensured that the stimulated electrical response caused by different OPAs could not have been simply due to variations in optical power between the OPAs. The recovery period between trials of the same optical input power but different OPAs was typically 20-60 s. The recovery period between trials where the input power was changed was 10-70 s and limited by the time required to adjust the variable neutral density filter.

\section{Spike Sorting of Microelectrode Array Recordings}

Spike sorting of the microelectrode array recordings was performed to analyze the data from the optogenetic stimulation experiments (Figs. 5 and S4). The full electrical traces from each experiment, spanning all optical pulse trains and the recovery periods between them, were analyzed. We performed spike sorting with the Spyking Circus package \cite{yger2018spike}. We selected electrode channels where neuronal activity was detected in response to the optical stimulation. A bandpass filter with a passband from 300 to 3000 Hz was applied to the signal from each electrode channel. Then, any negative spike with amplitude larger than 6 times the mean absolute deviation was selected as a valid spike. Each spike waveform was extracted from a 3 ms time window centered at the spike peak. Then, to extract spike templates, a subset of 10000 spikes was selected; for data sets with $< 10000$ spikes, all spikes were selected. The selected spikes were projected to a lower dimension using Principal Component Analysis. We selected the first 5 prominent components as the basis. Then a density-based spike clustering algorithm was applied to cluster spikes with similar waveforms \cite{yger2018spike}. The median of the spike waveform in the same cluster was defined as the spike template of the cluster. Template matching was performed to decompose all spikes as a linear combination of the templates, addressing the problem of overlapping spikes \cite{yger2018spike}. Lastly, we performed a manual inspection of the sorted spikes in the phy GUI interface \cite{cyrillerossant}. We selected the clusters that meet the following four criteria: 1) isolation distance $> 10$, 2) likelihood ratio $< 0.3$, 3) signal to noise ratio (SNR) $> 2$ \cite{schmitzer2005quantitative,libbrecht2018proximal,rutishauser2006online}, and 4) the percentage of spikes with interspike intervals $< 2$ ms was less than 2\% when only considering the spikes in the optical stimulation windows. Only sorted spikes within an optical stimulation pulse window were accounted for in the spike rate calculations.

\section{Supplementary Figures}

\begin{figure*}[ht!]
    \centering
    \includegraphics[width=4.7in]{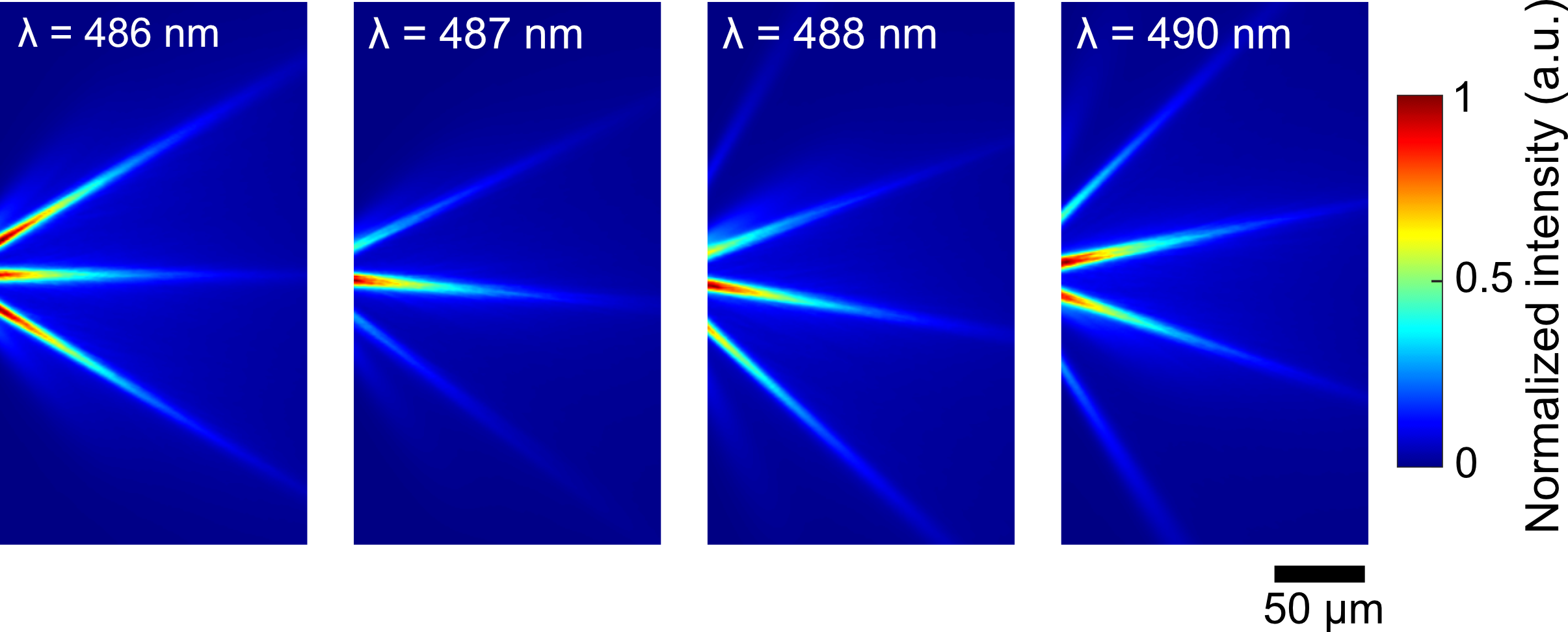}
    \caption{Additional data for the OPA top-down beam profile simulations in brain tissue shown in Fig. 3(a). Here, the beam profiles from Fig. 3(a) have been cropped by 30 $\mu m$ on the left side (closest to the OPA). The visibility of the beam profiles is improved since the large beam intensity at the OPA emitter is removed and each intensity color scale is normalized to the maximum of the lobes after they have separated.}
    \label{fig:FigureS1}
\end{figure*}


\begin{figure*}[ht!]
    \centering
    \includegraphics[width=5.1in]{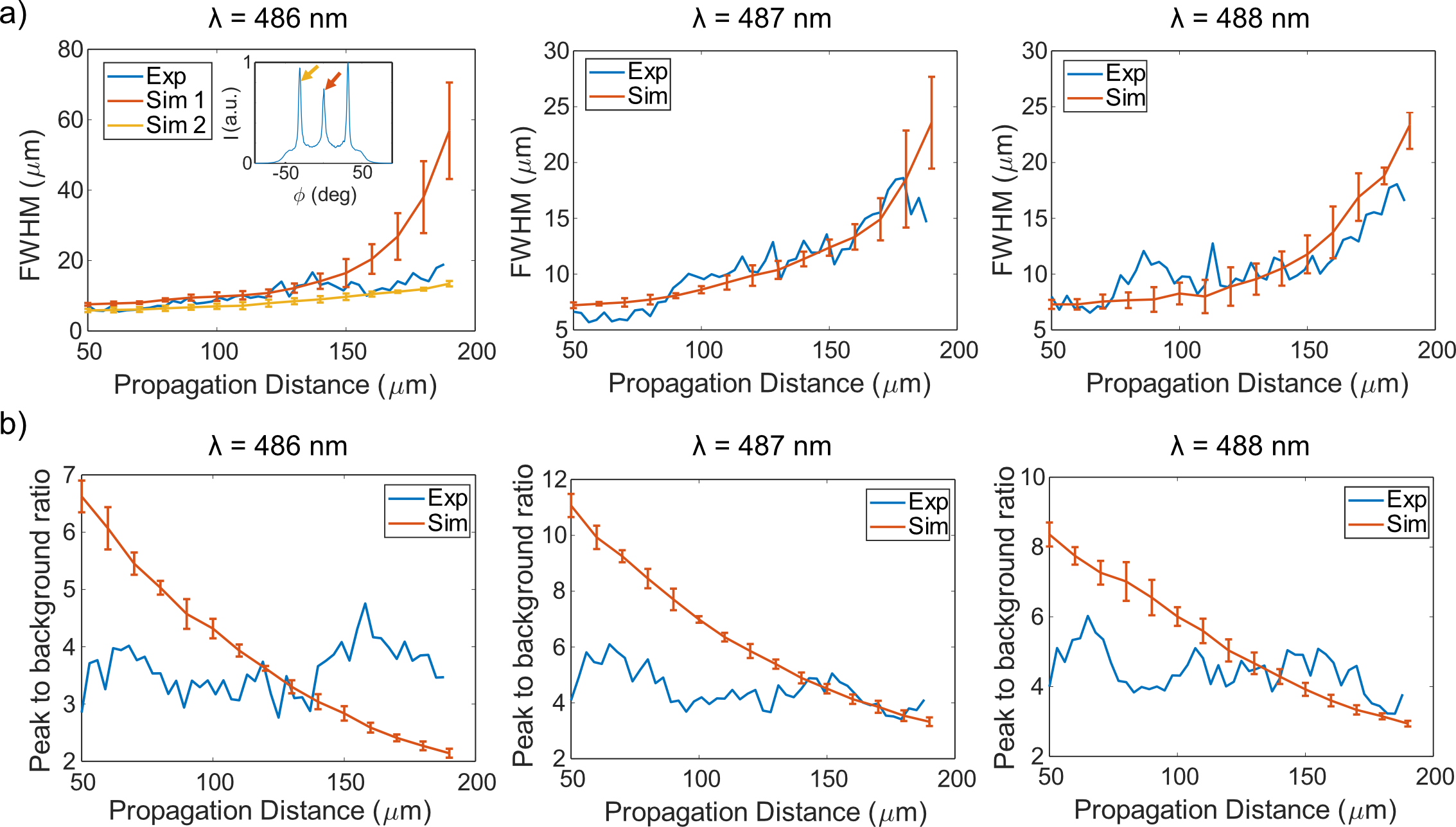}
    \caption{Additional data for the OPA top-down beam profile simulations and measurements shown in Fig. 3. Simulated and measured intensity beam profile characteristics are shown at wavelengths of 486, 487 and 488 nm. (a) Measured and simulated FWHM beam width versus propagation distance. (b) Measured and simulated ratios of peak beam intensity to background intensity versus propagation distance. Each data point for the simulated beam profiles from (a) and (b) is an average over 5 simulations with different random phase masks. The bars represent the standard deviation of the 5 simulations at each propagation distance. The FWHM plot for $\lambda = 486$ nm shows 2 simulations: "Sim 1" is the simulated FWHM of the central lobe emitted by the OPA, and "Sim 2" is the FWHM of a side lobe. The inset shows the simulated intensity ($I$) profile from Fig. 3(a) (propagation distance $= 100$ $\mu$m, $\lambda = 486$ nm) with arrows indicating the lobes corresponding to "Sim 1" and "Sim 2". The central lobe is less intense than the side lobes at this wavelength, a result of the non-uniformity of the emissions of the individual gratings  in the OPA versus steering angle $\phi$, i.e., the envelope of the emitters. Consequently, for propagation distances beyond $\approx 120$ $\mu$m, the background intensity contributes more significantly to the FWHM of the central lobe than the side lobes, leading to a large FWHM for "Sim 1" relative to "Sim 2" and $\lambda = $ 487 and 488 nm.}
    \label{fig:FigureS2}
\end{figure*}

\begin{figure*}[ht!]
    \centering
    \includegraphics[width=2.5in]{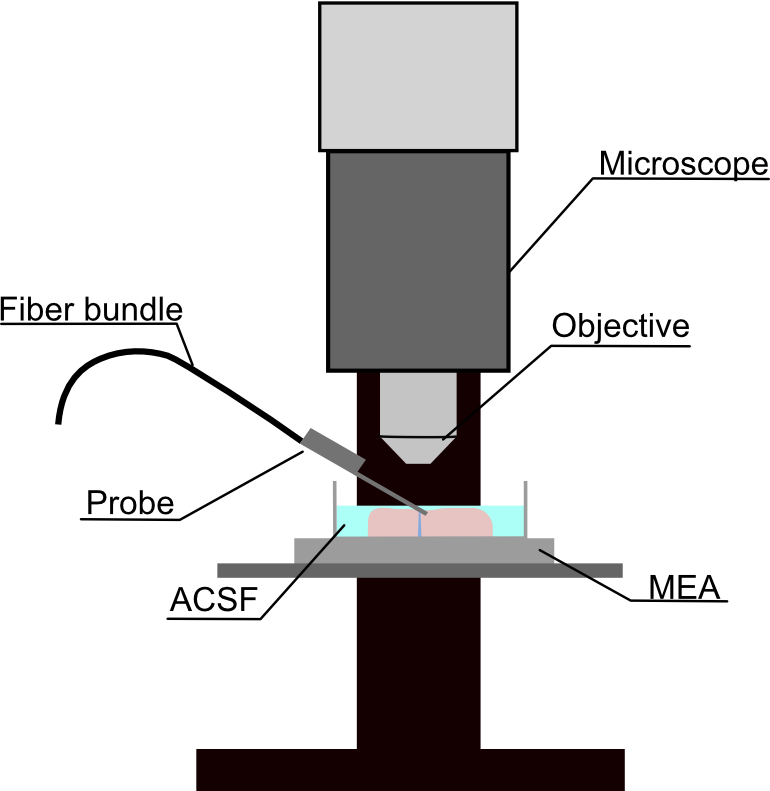}
    \caption{Illustration of the experimental apparatus for testing optogenetic stimulation of brain slices using the OPA neural probe. Here, the probe has a different orientation compared to Fig. 1(d). This experimental apparatus was used for the measurements in Figs. 5 and S4.}
    \label{fig:FigureS3}
\end{figure*}

\bigskip
\bigskip

\begin{figure*}[ht!]
    \centering
    \includegraphics[width=4.8in]{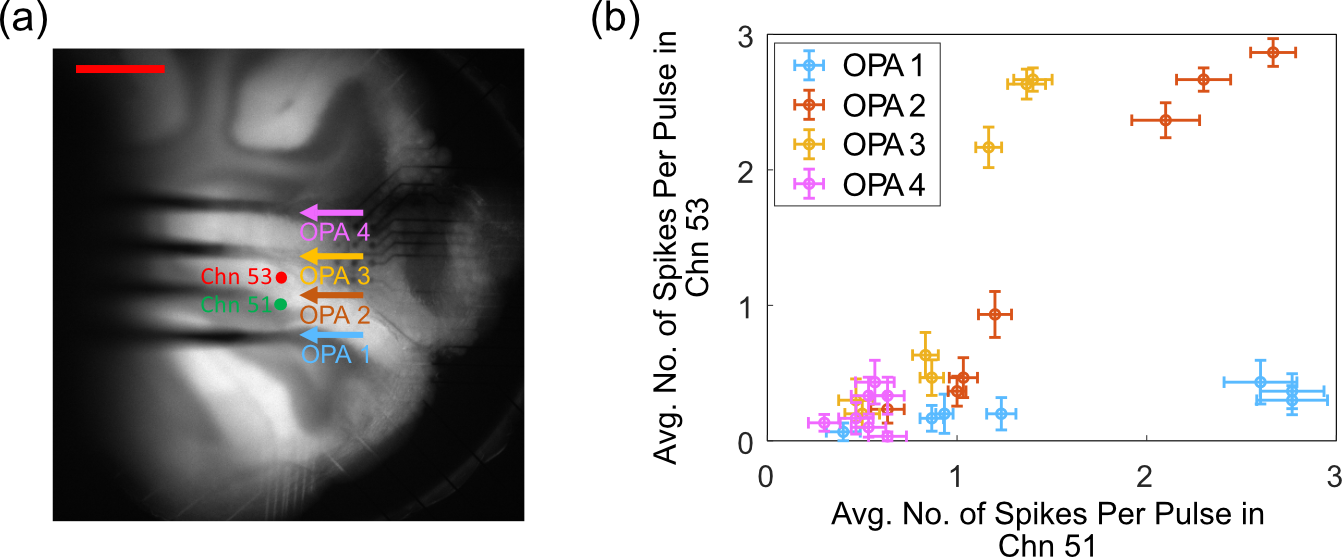}
    \caption{Additional optogenetic stimulation experiment showing the stimulation effects of 4 of the neural probe OPAs on an \emph{in vitro} brain slice. The experiment was performed on a 350 $\mu$m thick sagittal cerebellar slice obtained from a VGAT-ChR2-EYFP mice; the same brain slice as in Figs. 5(c) and (d). The fluorescence image (a) shows the relative positions of the OPAs with respect to the MEA channels of interest; epi-illumination from the top-down microscope was applied for visibility of the shanks and the structure of the brain slice. The scale bar is 500 $\mu$m. (b) shows the average number of spikes per optical pulse from 4 of the OPAs on two different MEA channels (``Chn''). As in Fig. 5(d), each point in (b) is from one stimulation trial consisting of 3 sets of 10 optical pulses; the optical pulse width was 50 ms, the period was 200 ms, and the recovery period between each set of 10 optical pulses was 10 s. The optical power was reduced across the trials for each OPA. The error bars represent the standard error of the mean number of spikes per pulse in a trial.}
    \label{fig:FigureS4}
\end{figure*}

\end{document}